\documentstyle[prb,aps]{revtex}
\begin{document}

\title{Energy Levels of Multielectron Atoms using the Sea-boson Method}
\author{Girish S. Setlur and Yia-Chung Chang}
\address{Department of Physics and Materials Research Laboratory,\\
 University of Illinois at Urbana-Champaign , Urbana Il 61801}
\maketitle

\begin{abstract}
 In this article, we study an antithesis of the homogeneous Fermi system
 (jellium model) namely a single multielectron atom using the
 sea-boson approach. This system in addition
 to having a small number of electrons(finite system) is also extremely 
 inhomogeneous. This means that the plane-wave basis is ill-suited to study
 this system, instead we use a basis consisting of the eigenfunctions 
 of an electron interacting with the nucleus situated at the origin
 (ideal atom).
 We compute the energy levels of
 this system and polarizabilities and compare with the results of other
 approaches such as Hartree-Fock and pseudopotentials for the former
 and aysmptotic wavefunctions for the latter and with experiments.
\end{abstract}

\section{Introduction}

 In our previous work\cite{Setlur1} we showed how to compute the
 single-particle properties of a homogeneous
 Fermi system using the sea-boson method.
 We also generalised these ideas
 to include spin into the formalism\cite{Setlur2}. 
 However, in all these approaches the implicit assumption was that the
 plane-wave basis was the right one to study the system at hand. 
 While this assumption is no doubt an excellent one for translationally
 invariant systems and perhaps even for systems that are nearly
 homogeneous such as a crystal, it is extremely ill-suited to study 
 very inhomogeneous systems such as organic substances or to take an
 extreme example like the one to be studied in this article, a single atom. 
 For a single atom, we have to choose a basis that properly reflects the
 highly localised nature of the distribution of electrons in real space.
 For this reason we find that it is best not to use the plane wave basis
 but its antithesis, namely  wavefunctions that are eigenfunctions of the
 noninteracting electrons interacting with just the nucleus.
 In the next section, we show how to write
 down formulas for the number conserving product of two Fermi/Bose fields
 in terms of the sea-bosons but this time in the localised
 (relatively speaking, relative to plane-waves) basis
 which are the eigenfunctions of the electrons interacting with the 
 nucleus at the origin. 
 Following this, we write down a model hamiltonian that describes a system
 consisting of electrons interacting with themelves and with a bosonic
 nucleus situated at the origin. This may be diagonalised exactly
 in what is now an analog of RPA in the localised basis and go on to compute
 the energy levels of this system. 

\section{Parent-Bilinear Sea-Boson Correspondence}

 Let us define $ H_{0} = \frac{-\nabla^{2}}{2m} + U({\vec{r}}) $,
 $ U({\vec{r}}) = -Z\mbox{    }e^{2}/r $.
 Then its eigenfunctions may be written down as follows,
\begin{equation}
H_{0} \varphi_{i}({\vec{r}}) = \epsilon_{i} \varphi_{i}({\vec{r}}) 
\end{equation}
The hamiltonian for this problem may be cast in the following form,
\begin{equation}
K = \int \mbox{      }d{\vec{r}}\mbox{       }
\sum_{ \sigma } 
\psi^{\dagger}({\vec{r}}\sigma)
(\frac{-\nabla^{2}}{2m} + U({\vec{r}}))
\psi({\vec{r}}\sigma)
\end{equation}
This may be written in the canonical basis as,
\begin{equation}
K = \sum_{ {\vec{i}} \sigma }\epsilon_{ {\vec{i}} }
c^{\dagger}({\vec{i}}\sigma)c({\vec{i}}\sigma)
\label{HAMILT0}
\end{equation}
The seaboson is defined as follows,
\begin{equation}
a_{ {\vec{i}}\sigma }({\vec{i^{'}}}\sigma^{'})
 = \frac{1}{\sqrt{n({\vec{i}}\sigma)}}
c^{\dagger}({\vec{i}}\sigma)
(\frac{ {\bar{n}}^{0}({\vec{i}}\sigma) }{ \langle N \rangle })^{\frac{1}{2}}
e^{i\vartheta({\vec{i}}\sigma,{\vec{i^{'}}}\sigma^{'})}
c({\vec{i^{'}}}\sigma^{'}); \mbox{       }{\vec{i}} \neq {\vec{i^{'}}}
\label{DEFN}
\end{equation}
\begin{equation}
a_{ {\vec{i}}\sigma }({\vec{i}}\sigma^{'})
 = 0
\end{equation}
here,
\begin{equation}
\langle N \rangle  = \sum_{ {\vec{i}}\sigma }
{\bar{n}}^{0}({\vec{i}}\sigma)
\label{CONSTR}
\end{equation}
Here $ {\bar{n}}^{0}({\vec{i}}\sigma) $ is the ground state distribution of
 electrons of the ideal atom.
 For example in the case of helium $ {\bar{n}}^{0}({\vec{i}}=1s, \sigma) =1 $,
 for either spin, and $ {\bar{n}}^{0}({\vec{i}}=2s,2p, ... \sigma) = 0 $.
\[
c^{\dagger}({\vec{i}}\sigma)c({\vec{i^{'}}}\sigma^{'})
 = \Lambda_{ {\vec{i}}\sigma }({\vec{i^{'}}}\sigma^{'})
a_{ {\vec{i}}\sigma }({\vec{i^{'}}}\sigma^{'})
 + \Lambda_{ {\vec{i^{'}}}\sigma^{'} }({\vec{i}}\sigma)
a^{\dagger}_{ {\vec{i^{'}}}\sigma^{'} }({\vec{i}}\sigma)
\]
\begin{equation}
+ \sqrt{1- {\bar{n}}({\vec{i}}\sigma)}
\sqrt{1- {\bar{n}}({\vec{i^{'}}}\sigma^{'})}
\sum_{ {\bf{i}}_{1} \sigma_{1} }
a^{\dagger}_{ {\bf{i}}_{1} \sigma_{1} }({\vec{i}}\sigma)
a_{ {\bf{i}}_{1} \sigma_{1} }({\vec{i^{'}}}\sigma^{'})
 - \sqrt{{\bar{n}}({\vec{i}}\sigma)}
\sqrt{{\bar{n}}({\vec{i^{'}}}\sigma^{'})}
\sum_{ {\bf{i}}_{1} \sigma_{1} }
a^{\dagger}_{ {\bf{i^{'}}} \sigma^{'} }({\vec{i}}_{1}\sigma_{1})
a_{ {\bf{i}} \sigma }({\vec{i}}_{1}\sigma_{1})
\label{CORR}
\end{equation}
The density operator has a similar formula,
\begin{equation}
n({\vec{i}}\sigma) = 
 c^{\dagger}({\vec{i}}\sigma)c({\vec{i}}\sigma)
 = {\bar{n}}^{0}({\vec{i}}\sigma)\frac{ N }{ \langle N \rangle }
 + \sum_{ {\bf{i}}_{1} \sigma_{1} }
a^{\dagger}_{ {\bf{i}}_{1} \sigma_{1} }({\vec{i}}\sigma)
a_{ {\bf{i}}_{1} \sigma_{1} }({\vec{i}}\sigma)
- \sum_{ {\bf{i}}_{1} \sigma_{1} }
a^{\dagger}_{ {\bf{i}} \sigma }({\vec{i}}_{1}\sigma_{1})
a_{ {\bf{i}}\sigma }({\vec{i}}_{1}\sigma_{1})
\label{NUMBER}
\end{equation}
 also, $ N_{\sigma} = \sum_{ {\vec{i}} }n({\vec{i}}\sigma) $.
Furthermore, the sea-bosons are canonical bosons,
\begin{equation}
[a_{ {\vec{i}}\sigma }({\vec{i^{'}}}\sigma^{'}), 
a_{ {\vec{j}}\rho }({\vec{j^{'}}}\rho^{'})] = 0
\end{equation}
\begin{equation}
[a_{ {\vec{i}}\sigma }({\vec{i^{'}}}\sigma^{'}),
a^{\dagger}_{ {\vec{j}}\rho }({\vec{j^{'}}}\rho^{'})] = 
\delta_{ {\vec{i}}\sigma, {\vec{j}}\rho }
\delta_{ {\vec{i^{'}}}\sigma^{'}, {\vec{j^{'}}}\rho^{'} }
\label{COMM}
\end{equation}
 The above correspondence reproduces the following salient features of the
 free theory,
\newline
(1) The definition in Eq.(~\ref{DEFN}) when plugged into the formula for
 the number operator Eq.(~\ref{NUMBER}) gives an identity.
\newline
(2) The dynamical four-point and six-point functions of the free
 theory are reproduced properly. That is if one computes the
 correlators of the fermi bilinears in Eq.(~\ref{CORR}) using the free
 hamiltonian in Eq.(~\ref{HAMILT0}), one gets what one expects.
\newline
(3) The commutation rules are reproduced only in the "RPA"-sense.

%

\section{The Hamiltonian of the Multi-electron Atom}

Let us write down the hamiltonain of the multi-electron atom
 ($ H = K + U $),
\begin{equation}
K = \int {\mbox{       }} d{\vec{r}}  {\mbox{       }}
\sum_{ \sigma }
\psi^{\dagger}({\vec{r}}\sigma)(-\frac{\nabla^{2}}{2m} + U({\vec{r}}))
\psi({\vec{r}}\sigma)
\end{equation}
\begin{equation}
U =  
 \frac{1}{2}\int {\mbox{       }} d{\vec{r}}  {\mbox{       }}
\int {\mbox{       }} d{\vec{r^{'}}}  {\mbox{       }}
v({\vec{r}}-{\vec{r^{'}}})
\sum_{ \sigma, \sigma^{'} }
\psi^{\dagger}({\vec{r}}\sigma)\psi^{\dagger}({\vec{r^{'}}}\sigma^{'})
\psi({\vec{r^{'}}}\sigma^{'})\psi({\vec{r}}\sigma)
\end{equation}
Here $  U({\vec{r}}) = -Z\mbox{   }e^{2}/r $
and $ v({\vec{r}}-{\vec{r^{'}}}) = e^{2}/|{\vec{r}}-{\vec{r^{'}}}| $.
\begin{equation}
K = E_{0} + \sum_{{\vec{i}},{\vec{i}}_{1},\sigma,\sigma_{1}}
(\epsilon_{i}-\epsilon_{i_{1}})a^{\dagger}_{{\vec{i}}_{1}\sigma_{1}}
({\vec{i}}\sigma) a_{{\vec{i}}_{1}\sigma_{1}}({\vec{i}}\sigma)
\end{equation}
Similarly, the potential term $ U $ may be written as,
\begin{equation}
U =
 \frac{1}{2}\int {\mbox{       }} d{\vec{r}}  {\mbox{       }}
\int {\mbox{       }} d{\vec{r^{'}}}  {\mbox{       }}
v({\vec{r}}-{\vec{r^{'}}})
\sum_{ \sigma, \sigma^{'} }
\psi^{\dagger}({\vec{r}}\sigma)\psi^{\dagger}({\vec{r^{'}}}\sigma^{'})
\psi({\vec{r^{'}}}\sigma^{'})\psi({\vec{r}}\sigma)
\end{equation}
\begin{equation}
U =
\frac{1}{2}\sum_{ {\vec{i}}, {\vec{i^{'}}}, {\vec{j}}, {\vec{j^{'}}},
\sigma, \sigma^{'} }
 V_{ {\vec{i}}{\vec{j}}, {\vec{i^{'}}}{\vec{j^{'}}} }
c^{\dagger}({\vec{i}}\sigma)c^{\dagger}({\vec{i^{'}}}\sigma^{'})
c({\vec{j^{'}}}\sigma^{'})c({\vec{j}}\sigma)
\end{equation}
\begin{equation}
V_{ {\vec{i}}{\vec{j}}, {\vec{i^{'}}}{\vec{j^{'}}} } = 
  \int {\mbox{       }} d{\vec{r}}  {\mbox{       }}
\int {\mbox{       }} d{\vec{r^{'}}}  {\mbox{       }}
v({\vec{r}}-{\vec{r^{'}}})
\varphi_{{\vec{i}}}^{*}({\vec{r}})
\varphi^{*}_{{\vec{i^{'}}}}({\vec{r^{'}}})
\varphi_{{\vec{j}}}({\vec{r}})
\varphi_{{\vec{j^{'}}}}({\vec{r^{'}}})
\end{equation}

\subsection{Properties of the Ideal Atom in the Sea-boson Language}

 In the previous section $ E_{0} $ was the ground state energy. 
 It is computed as follows. Let us assume we have N electrons 
 in our system. This means that we have our eigenstates filled
 upto $ i_{max} $ starting from $ i = 0 $
 $ {\mathcal{B}} = \{ i = 0, 1, 2, ..., i_{max} \} $.
 each $ i $ has at most two electrons with spin up or down.
 Let us consider an even number of electrons,
\begin{equation}
 N = 2 \times (i_{max}+1)
\end{equation}
\begin{equation}
E_{0} = 2\epsilon_{0} + 2\epsilon_{1} + ... + 2\epsilon_{ i_{max} }
\end{equation}
The ground state of this system is annhilated by the sea-bosons,
\begin{equation}
a_{ {\vec{i}}\sigma }({\vec{j}}\sigma^{'}) |G\rangle = 0
\end{equation}
also the sea-boson has the propety that(a postulate if you like),
\begin{equation}
a_{ {\vec{i}}\sigma }({\vec{j}}\sigma^{'}) = 0 ; \mbox{          }
for  \mbox{           }\epsilon_{j} < \epsilon_{i}
\end{equation}
This allows us to construct the first excited state of this system as
\begin{equation}
|E_{1}\rangle = a^{\dagger}_{ i=i_{max}\sigma }( (j=i_{max}+1)\sigma^{'})
|G\rangle
\end{equation}
 where the spins $ \sigma, \sigma^{'} $ can be anything. It says that
 the first excited state is degenerate. That is, you can take a spin up or
 down electron situated at $ i = i_{max} $ and elevate it to 
 $ i = i_{max} + 1 $ and that electron can be of either spin as well.
 Hence the four-fold degeneracy.
 The energy of the first excited state is given by,
\begin{equation}
E_{1} =  2\epsilon_{0} + 2\epsilon_{1} + ... + 2\epsilon_{ i_{max} }
+ (\epsilon_{ i_{max}+1 } - \epsilon_{ i_{max} })
\end{equation}
 Having convinced ourselves of the correctness of the formalism as regards
 the ideal atom we proceed to study the full problem.

\section{The Nonideal Atom}

 In order to compute the energy levels of the nonideal atom, we proceed
 as follows. First let us rewrite the interaction term in the 
 sea-boson language. 
\begin{equation}
U =
\frac{1}{2}\sum_{ {\vec{i}}, {\vec{i^{'}}}, {\vec{j}}, {\vec{j^{'}}},
\sigma, \sigma^{'} }
 V_{ {\vec{i}}{\vec{j}}, {\vec{i^{'}}}{\vec{j^{'}}} }
c^{\dagger}({\vec{i}}\sigma)c({\vec{j}}\sigma)
c^{\dagger}({\vec{i^{'}}}\sigma^{'})
c({\vec{j^{'}}}\sigma^{'})
-N\mbox{     }V_{0}
\end{equation}
\begin{equation}
V_{0} = \frac{1}{2}
\sum_{ {\vec{j}} }V_{ {\vec{i}}{\vec{j}}, {\vec{j}}{\vec{i}} }
\end{equation}
\begin{equation}
U = \frac{1}{2}\sum_{ {\vec{i}}, {\vec{i^{'}}}, {\vec{j}}, {\vec{j^{'}}},
\sigma, \sigma^{'} }
 V_{ {\vec{i}}{\vec{j}}, {\vec{i^{'}}}{\vec{j^{'}}} }
[\Lambda_{ {\vec{i}}\sigma }({\vec{j}}\sigma)a_{ {\vec{i}}\sigma }
({\vec{j}}\sigma)
 + \Lambda_{ {\vec{j}}\sigma }({\vec{i}}\sigma)
a^{\dagger}_{ {\vec{j}}\sigma }({\vec{i}}\sigma)]
[\Lambda_{ {\vec{i^{'}}}\sigma^{'} }({\vec{j^{'}}}\sigma^{'})
a_{ {\vec{i^{'}}}\sigma^{'} }
({\vec{j^{'}}}\sigma^{'})
 + \Lambda_{ {\vec{j^{'}}}\sigma^{'} }({\vec{i^{'}}}\sigma^{'})
a^{\dagger}_{ {\vec{j^{'}}}\sigma^{'} }({\vec{i^{'}}}\sigma^{'})]
\end{equation}
Let us now diagonalise this hamiltonian.
For this we have to introduce the dressed sea-bosons.
\begin{equation}
d_{I\sigma} = \sum_{ {\vec{i}},{\vec{j}},\sigma,\sigma^{'}}
 [d_{I\sigma},a^{\dagger}_{ {\vec{i}}\sigma }({\vec{j}}\sigma^{'})]
a_{ {\vec{i}}\sigma }({\vec{j}}\sigma^{'})
 - \sum_{ {\vec{i}},{\vec{j}},\sigma,\sigma^{'}}
[d_{I\sigma},a_{ {\vec{i}}\sigma }({\vec{j}}\sigma^{'})]
a^{\dagger}_{ {\vec{i}}\sigma }({\vec{j}}\sigma^{'})
\end{equation}
Let the diagonalised hamiltonian be written as,
\begin{equation}
H = \sum_{I\sigma}\omega_{I\sigma}d^{\dagger}_{I\sigma}d_{I\sigma}
\end{equation}
\[
\omega_{I\sigma}d_{I\sigma} = \sum_{ {\vec{i}}, {\vec{j}}, \sigma, \sigma^{'} }
(\epsilon_{j}- \epsilon_{i})[d_{I\sigma},a^{\dagger}_{ {\vec{i}}\sigma }
({\vec{j}}\sigma^{'})]a_{ {\vec{i}}\sigma }({\vec{j}}\sigma^{'})
 +  \sum_{ {\vec{i}}, {\vec{j}}, \sigma, \sigma^{'} }
(\epsilon_{j}- \epsilon_{i})[d_{I\sigma},a_{ {\vec{i}}\sigma }
({\vec{j}}\sigma^{'})]a^{\dagger}_{ {\vec{i}}\sigma }({\vec{j}}\sigma^{'})
\]
\[
+ \frac{1}{2}
\sum_{ {\vec{i}}, {\vec{i^{'}}},{\vec{j}}, {\vec{j^{'}}},\sigma, \sigma^{'} }
V_{ {\vec{i}}{\vec{j}}, {\vec{i^{'}}}{\vec{j^{'}}} }
[\Lambda_{ {\vec{i}} \sigma }({\vec{j}}\sigma)
[d_{I\sigma},a_{ {\vec{i}} \sigma }({\vec{j}}\sigma)]
+ \Lambda_{ {\vec{j}} \sigma }({\vec{i}}\sigma)
[d_{I\sigma},a^{\dagger}_{ {\vec{j}} \sigma }({\vec{i}}\sigma)]]
[\Lambda_{ {\vec{i^{'}}} \sigma^{'} }({\vec{j^{'}}}\sigma^{'})
a_{ {\vec{i^{'}}} \sigma^{'} }({\vec{j^{'}}}\sigma^{'})
+ a^{\dagger}_{ {\vec{j^{'}}} \sigma^{'} }({\vec{i^{'}}}\sigma^{'})
\Lambda_{ {\vec{j^{'}}} \sigma^{'} }({\vec{i^{'}}}\sigma^{'})]
\]
\begin{equation}
+ \frac{1}{2}
\sum_{ {\vec{i}}, {\vec{i^{'}}},{\vec{j}}, {\vec{j^{'}}},\sigma, \sigma^{'} }
V_{ {\vec{i}}{\vec{j}}, {\vec{i^{'}}}{\vec{j^{'}}} }
[\Lambda_{ {\vec{i}} \sigma }({\vec{j}}\sigma)
a_{ {\vec{i}} \sigma }({\vec{j}}\sigma)
+ \Lambda_{ {\vec{j}} \sigma }({\vec{i}}\sigma)
a^{\dagger}_{ {\vec{j}} \sigma }({\vec{i}}\sigma)]
[\Lambda_{ {\vec{i^{'}}} \sigma^{'} }({\vec{j^{'}}}\sigma^{'})
[d_{I\sigma},a_{ {\vec{i^{'}}} \sigma^{'} }({\vec{j^{'}}}\sigma^{'})]
+ [d_{I\sigma},a^{\dagger}_{ {\vec{j^{'}}} \sigma^{'} }({\vec{i^{'}}}\sigma^{'})]
\Lambda_{ {\vec{j^{'}}} \sigma^{'} }({\vec{i^{'}}}\sigma^{'})]
\end{equation}
\begin{equation}
\omega_{I\sigma}[d_{I\sigma},a^{\dagger}_{ {\vec{i}}\sigma }({\vec{j}}\sigma^{'})] = 
(\epsilon_{j}- \epsilon_{i})[d_{I\sigma},a^{\dagger}_{ {\vec{i}}\sigma }
({\vec{j}}\sigma^{'})] + \sum_{ {\vec{i^{'}}}, {\vec{j^{'}}} }
V_{ {\vec{i}}{\vec{j}}, {\vec{i^{'}}}{\vec{j^{'}}} }
[\Lambda_{ {\vec{i^{'}}} \sigma }({\vec{j^{'}}}\sigma)
[d_{I\sigma},a_{ {\vec{i^{'}}} \sigma }({\vec{j^{'}}}\sigma)]
+ \Lambda_{ {\vec{j^{'}}} \sigma }({\vec{i^{'}}}\sigma)
[d_{I\sigma},a^{\dagger}_{ {\vec{j^{'}}} \sigma }({\vec{i^{'}}}\sigma)]]
\Lambda_{ {\vec{i}} \sigma }({\vec{j}}\sigma)
\delta_{ \sigma, \sigma^{'} }
\end{equation}
\begin{equation}
\omega_{I\sigma}[d_{I\sigma},a_{ {\vec{j}}\sigma^{'} }({\vec{i}}\sigma)] =
(\epsilon_{j} - \epsilon_{i})[d_{I\sigma},a_{ {\vec{j}}\sigma^{'} }
({\vec{i}}\sigma)] - \sum_{ {\vec{i^{'}}}, {\vec{j^{'}}} }
V_{ {\vec{i}}{\vec{j}}, {\vec{i^{'}}}{\vec{j^{'}}} }
[\Lambda_{ {\vec{i^{'}}} \sigma }({\vec{j^{'}}}\sigma)
[d_{I\sigma},a_{ {\vec{i^{'}}} \sigma }({\vec{j^{'}}}\sigma)]
+ \Lambda_{ {\vec{j^{'}}} \sigma }({\vec{i^{'}}}\sigma)
[d_{I\sigma},a^{\dagger}_{ {\vec{j^{'}}}\sigma }({\vec{i^{'}}}\sigma)]]
\Lambda_{  {\vec{j}} \sigma }({\vec{i}}\sigma)\delta_{ \sigma, \sigma^{'} }
\end{equation}
\begin{equation}
[d_{I\sigma},a^{\dagger}_{ {\vec{i}}\sigma }({\vec{j}}\sigma^{'})]
 = \frac{ \Lambda_{ {\vec{i}} \sigma }({\vec{j}}\sigma)
 \delta_{ \sigma, \sigma^{'} } }
{\omega_{I\sigma} - \epsilon_{j} + \epsilon_{i}}
G_{ \sigma }({\vec{i}}, {\vec{j}})
\end{equation}
\begin{equation}
[d_{I\sigma},a_{ {\vec{j}}\sigma^{'} }({\vec{i}}\sigma)]
 = -\frac{ \Lambda_{ {\vec{j}} \sigma }({\vec{i}}\sigma)
 \delta_{ \sigma, \sigma^{'} } }
{\omega_{I\sigma} - \epsilon_{j} + \epsilon_{i}}
G_{ \sigma }({\vec{i}}, {\vec{j}})
\end{equation}
The eigenvalues $ \omega_{I\sigma} $ that correspond to energies of the 
 excited states measured with respect to the ground states are given by the
 solution to the following eigenvalue problem,
\begin{equation}
G_{\sigma}({\vec{i^{'}}}, {\vec{j^{'}}})
 = \sum_{ {\vec{i}}, {\vec{j}} }
\frac{ {\bar{n}}_{ {\vec{i}}\sigma } -  {\bar{n}}_{ {\vec{j}}\sigma }  }
{\omega_{I\sigma} - \epsilon_{j} + \epsilon_{i}}
G_{\sigma}({\vec{i}}, {\vec{j}})
V_{ {\vec{i^{'}}}{\vec{j^{'}}}, {\vec{j}}{\vec{i}} }
\end{equation}
here we must impose the additional constraint,
\begin{equation}
G_{\sigma}({\vec{i}}, {\vec{j}}) = 0;\mbox{           }
\epsilon_{j} < \epsilon_{i}
\end{equation}
in order that excited states have higher energies than the ground state.
Further since we have constraint $ [d_{I\sigma}, d^{\dagger}_{I\sigma}] = 1 $,
 the matrix $ G_{\sigma} $ may itself be computed by the following equation,
\begin{equation}
1 = \sum_{i,j}\frac{ {\bar{n}}^{0}({i\sigma}) -  {\bar{n}}^{0}({j\sigma}) }
{(\omega_{I\sigma} - \epsilon_{j} + \epsilon_{i})^{2}}
G^{2}_{\sigma}(i,j)
\end{equation}
Also,
\begin{equation}
d_{I\sigma} = \sum_{i,j}[d_{I\sigma},a^{\dagger}_{i\sigma}(j\sigma)]
a_{i\sigma}(j\sigma) - \sum_{i,j}[d_{I\sigma},a_{j\sigma}(i\sigma)]
a^{\dagger}_{j\sigma}(i\sigma)
\end{equation}
The full hamiltonian may then be rewritten more transparently as,
\[
H = E^{'}_{0} + \sum_{ \epsilon_{j} > \epsilon_{i} }
(\epsilon_{j}-\epsilon_{i})
(a^{\dagger}_{ {\vec{i}}\uparrow }({\vec{j}}\downarrow)
a_{ {\vec{i}}\uparrow }({\vec{j}}\downarrow)
 + a^{\dagger}_{ {\vec{i}}\downarrow }({\vec{j}}\uparrow)
a_{ {\vec{i}}\downarrow }({\vec{j}}\uparrow) )
\]
\begin{equation}
+ \sum_{I}\omega_{I\uparrow}d^{\dagger}_{I\uparrow}d_{I\uparrow}
 + \sum_{I}\omega_{I\downarrow}d^{\dagger}_{I\downarrow}d_{I\downarrow}
\end{equation}
  From the above form of the full hamiltonian 
  it is clear that having obtained the eigenvalues
 $ \omega_{I\sigma} $ the next step would be to arrange the various excited 
 states according to
 whether or not $ (\epsilon_{j}-\epsilon_{i}) > \omega_{I\sigma} $
 There is also the issue of whether we are allowed to excite any number of
 sea-bosons all with the same label $ I\sigma $ (since they are bosons).
 The answer is no, since they have to describe particle-hole excitations
 of the atom and the elemetary particles are electrons.
 In order to answer this
 question we  have to follow closely the section where we showed how to 
 study the ideal atom in the sea-boson language. 
 There however we conveniently glossed
 over this thorny issue(sort of).
 But it is hoped that the reader appreciates the spirit
 of the argument and will not examine this flaw under a microscope. 
 The whole formalism has to be taken with a grain of salt 
 (a tall glass of milk, a jar of pickles and a can of diet coke !).

 The ground state energy of the system may be obtained by merely computing
 the expectation value of the kinetic energy
 and the potential energy separately and adding them together. 
\[
E^{'}_{0} = \sum_{ {\vec{i}} \sigma }\epsilon_{ {\vec{i}} }
{\bar{n}}^{0}( {\vec{i}}\sigma )
 + \sum_{ {\vec{i}},{\vec{j}} }
(\epsilon_{ {\vec{j}} } - \epsilon_{ {\vec{i}} })
(\langle a^{\dagger}_{ {\vec{i}}\uparrow }({\vec{j}}\uparrow)
 a_{ {\vec{i}}\uparrow }({\vec{j}}\uparrow) \rangle
 + \langle a^{\dagger}_{ {\vec{i}}\downarrow }({\vec{j}}\downarrow)
 a_{ {\vec{i}}\downarrow }({\vec{j}}\downarrow) \rangle)
\]
\[
+ \frac{1}{2}\sum_{ {\vec{i}}{\vec{j}},{\vec{i}}^{'}{\vec{j}}^{'} }
\sum_{ \sigma }
V_{ {\vec{i}}{\vec{j}},{\vec{i}}^{'}{\vec{j}}^{'} }
(\Lambda_{ {\vec{i}}\sigma }({\vec{j}}\sigma)
\Lambda_{ {\vec{i^{'}}}\sigma }({\vec{j^{'}}}\sigma)
\langle a_{ {\vec{i}}\sigma }({\vec{j}}\sigma)
a_{ {\vec{i^{'}}}\sigma }({\vec{j^{'}}}\sigma) \rangle
 + \Lambda_{ {\vec{i}}\sigma }({\vec{j}}\sigma)
\Lambda_{ {\vec{j^{'}}}\sigma }({\vec{i^{'}}}\sigma)
\langle a_{ {\vec{i}}\sigma }({\vec{j}}\sigma) 
a^{\dagger}_{ {\vec{j^{'}}}\sigma }({\vec{i^{'}}}\sigma) \rangle
\]
\begin{equation}
+ \Lambda_{ {\vec{j}}\sigma }({\vec{i}}\sigma)
\Lambda_{ {\vec{i^{'}}}\sigma }({\vec{j^{'}}}\sigma)
\langle a^{\dagger}_{ {\vec{j}}\sigma }({\vec{i}}\sigma)
a_{ {\vec{i^{'}}}\sigma }({\vec{j^{'}}}\sigma) \rangle
 + \Lambda_{ {\vec{j}}\sigma }({\vec{i}}\sigma)
\Lambda_{ {\vec{j^{'}}}\sigma }({\vec{i^{'}}}\sigma)
\langle a^{\dagger}_{ {\vec{j}}\sigma }({\vec{i}}\sigma)
a^{\dagger}_{ {\vec{j^{'}}}\sigma }({\vec{i^{'}}}\sigma) \rangle)
 - \langle N \rangle V_{0}
\end{equation}
Since,
\begin{equation}
a_{ {\vec{i}} \sigma }({\vec{j}}\sigma)
 = \sum_{ I }[a_{ {\vec{i}} \sigma }({\vec{j}}\sigma),d^{\dagger}_{ I \sigma }]
d_{ I \sigma }
 - \sum_{ I }[a_{ {\vec{i}} \sigma }({\vec{j}}\sigma),d_{ I \sigma }]
d^{\dagger}_{ I \sigma }
\end{equation}
\begin{equation}
a^{\dagger}_{ {\vec{i}} \sigma }({\vec{j}}\sigma)
 = \sum_{ I }[a_{ {\vec{i}} \sigma }({\vec{j}}\sigma),d^{\dagger}_{ I \sigma }]
d^{\dagger}_{ I \sigma }
 - \sum_{ I }[a_{ {\vec{i}} \sigma }({\vec{j}}\sigma),d_{ I \sigma }]
d_{ I \sigma }
\end{equation}
From these we may deduce,
\begin{equation}
\langle a_{ {\vec{i}} \sigma }({\vec{j}}\sigma)
 a_{ {\vec{i^{'}}} \sigma }({\vec{j^{'}}}\sigma) \rangle
 = -\sum_{ I }[a_{ {\vec{i}} \sigma }({\vec{j}}\sigma),d^{\dagger}_{ I \sigma }]
[a_{ {\vec{i^{'}}} \sigma }({\vec{j^{'}}}\sigma),d_{ I \sigma }]
\end{equation}
\begin{equation}
\langle a^{\dagger}_{ {\vec{i}} \sigma }({\vec{j}}\sigma)
 a^{\dagger}_{ {\vec{i^{'}}} \sigma }({\vec{j^{'}}}\sigma) \rangle
 = -\sum_{ I }[a_{ {\vec{i^{'}}} \sigma }({\vec{j^{'}}}\sigma)
,d^{\dagger}_{ I \sigma }]
[a_{ {\vec{i}} \sigma }({\vec{j}}\sigma),d_{ I \sigma }]
\end{equation}
\begin{equation}
\langle a^{\dagger}_{ {\vec{i}} \sigma }({\vec{j}}\sigma)
a_{ {\vec{i^{'}}} \sigma }({\vec{j^{'}}}\sigma) \rangle
 = \sum_{ I }[a_{ {\vec{i}} \sigma }({\vec{j}}\sigma),d_{ I \sigma }]
[a_{ {\vec{i^{'}}} \sigma }({\vec{j^{'}}}\sigma),d_{ I \sigma }]
\end{equation}
\begin{equation}
\langle a_{ {\vec{i}} \sigma }({\vec{j}}\sigma)
a^{\dagger}_{ {\vec{i^{'}}} \sigma }({\vec{j^{'}}}\sigma) \rangle
 = \sum_{ I }[a_{ {\vec{i}} \sigma }({\vec{j}}\sigma),d^{\dagger}_{ I \sigma }]
[a_{ {\vec{i^{'}}} \sigma }({\vec{j^{'}}}\sigma),d^{\dagger}_{ I \sigma }]
\end{equation}
The rest are computational details and will be done the next time we replace
 this preprint on the Los Alamos Archive.


\begin{thebibliography}{6}

\bibitem[1]{Setlur1} G.S. Setlur and Y.C. Chang, Phys. Rev. B15, June 15,
 vol 57, no. 24, 15 144(1998)

\bibitem[2]{Setlur2} G.S. Setlur and Y.C. Chang, cond-mat/9808264

\bibitem[4]{Hartree}See for example, {\it{Quantum Mechanics}} A.A. Sokolov,
 I.M. Ternov and V.Ch.Zhukovskii, Mir Publishers, Moscow \copyright 1984.

\bibitem[5]{Pseudo}W. Kutzelnigg and F. Maeder, Chem. Phys. {\bf{32}}, 451 
(1978); {\bf{35}}, 397 (1978); {\bf{42}}, 95 (1979)

\bibitem[6]{Setlur3} S.H. Patil and G.S. Setlur,
 J. Chem. Phys. {\bf{95}} (6) 4245-57 (1991)

\end{thebibliography}
\end{document}